\documentclass[11pt]{article}
\usepackage{moriond,epsfig}

\begin{document}
\vspace*{4cm}
\title{RECENT RESULTS IN THE SEARCH FOR DARK MATTER WITH NOBLE LIQUID DETECTORS \footnote{Presented at Rencontres de Moriond 2011 (Electroweak Session), La Thuile, Italy.}}

\author{A.~Manalaysay} 
\address{Physics Institute, University of Zurich, Winterthurerstrasse 190,\\
CH-8057 Z\"urich, Switzerland}

\maketitle\abstracts{
The field of dark matter direct detection has seen important contributions in recent years from experiments involving liquid noble gases, specifically liquid argon and liquid xenon.  These detection media offer many properties deemed useful in this search, including fast scintillation response, charge readout, 3-D position reconstruction, and nuclear recoil discrimination.  Part of the very rapid emergence and dominance of noble liquids is due to the fact that these technologies are easily scalable to nearly arbitrary size and mass.  However, the physics impact of recent results has called into question our understanding of the low-energy response of these detection media, in light of apparent contradictions with a possible low-mass WIMP signal observed in the DAMA/LIBRA and CoGeNT experiments.  I discuss recent results and examine the details of this inconsistency. 
}

\section{Introduction}
\label{sec:intro}

Stable, Weakly Interacting Massive Particles (WIMPs) naturally arise in a number of theories beyond the Standard Model of particle physics \cite{Bertone:2004pz}.  If nature allows for the existence of such a particle, it could have been produced thermally in the early Universe, resulting in a relic population persisting through to the present day, constituting what we observe as dark matter in astronomical observations \cite{Begeman:1991iy,Dunkley:2008ie,Clowe:2006eq}.  Under this scenario, the Earth is embedded within a gas of WIMPs having a characteristic local energy density of $\sim$0.3\,GeV/cm$^3$, a roughly Maxwell-Boltzman velocity distribution with characteristic velocities of $\mathcal{O}$(10$^{-3}\,\mathrm{c}$), and a weak interaction cross section.  Interactions between galactic WIMPs and atomic nuclei would produce roughly exponentially-falling differential energy spectra in terrestrial particle detectors, with energy depositions up to several---to several tens---of keV.  The exact details of the expected recoil spectra depend on the target nucleus and type of particles exchanged in the interaction, in addition to specific details of the astrophysical properties of the dark matter halo.  Typical interaction rates are expected to be low, ranging from a few counts/kg/day to a few counts/kg/year or fewer.  These low rates are in stark contrast to background rates in most particle detectors of $\mathcal{O}$(Hz) (from natural radioactivity and cosmic rays, for example), and therefore low-background techniques must be used to either reduce these backgrounds or otherwise distinguish signal from background. To drastically reduce the effect of cosmic rays, WIMP dark matter searches universally utilize underground laboratories, which provide factors of 10$^{-5}$ to 10$^{-8}$ reduction in the atmospheric muon flux.

Liquid argon (LAr) and liquid xenon (LXe), as particle detection media, have many properties that are beneficial from the standpoint of a low-background WIMP search.  Among the most important properties of these materials is the ability to design such a detector of almost arbitrary size.  A large detector has the ability to \emph{self shield} itself from backgrounds due to radioactive isotopes present in other detector materials.  This means that the outer regions of the detector can prevent much of these backgrounds from reaching the inner detector regions.  This extremely simple property has been shown to be immensely effective at reducing the overall background rate, as compared with technologies utilizing other detector materials.  Most background interactions arise from either gamma emitters in detector materials or beta emitters in the liquids themselves, whose energy deposition is characterized by recoiling electrons.  This is in contrast to the expected WIMP signal which is highly dominated by nuclear recoils.  Both LAr and LXe are able to reject electronic recoils ($\mathcal{O}(10^{-7})$ in LAr and $\mathcal{O}(10^{-3})$ in LXe).  

Additional benefits of liquid noble detectors include a high scintillation yield, 3-D position reconstruction capabilities, fast response of $\mathcal{O}$(few ns), ``easy'' cryogenics (compared with semiconductor detectors), high sensitivity to scalar interactions in LXe ($A\sim131$, scalar interaction rate $\propto A^2$), and sensitivity to axial vector interactions (48\% odd isotopes in LXe).

Past, current, and future WIMP searches using noble liquids include DAMA/LXe\,\cite{Bernabei:1998ad}, 
ZEPLIN-I\,\cite{Alner:2005pa}, 
ZEPLIN-II\,\cite{Alner:2007ja}, 
ZEPLIN-III\,\cite{Lebedenko:2008gb}, 
WArP 2.3l\,\cite{Benetti:2007cd}, 
WArP 100l\,\cite{WArP_100_confProc}, 
ArDM\,\cite{Haranczyk:2010cf}, 
DEAP/CLEAN\,\cite{DEAP_CLEAN_website}, 
DarkSide\,\cite{DarkSide_projSum}, 
XENON10\,\cite{Angle:2007uj}, 
XENON100\,\cite{Aprile:2010um}, 
XENON1t\,\cite{XENON1T_slides}, 
XMASS\,\cite{Sekiya:2010bf}, 
LUX\,\cite{Hall:2010zz}, 
LZ\,\cite{LZS_LZD_slides}, 
PANDA-X\,\cite{PANDAX_slides}, 
MAX\,\cite{Alarcon:2009zz}, and 
DARWIN\,\cite{Baudis:2010ch}.
Of these, seven have released dark matter results\,\cite{Bernabei:1998ad,Alner:2005pa,Alner:2007ja,Lebedenko:2008gb,Benetti:2007cd,Angle:2007uj,Aprile:2010um}.  I devote Section \ref{sec:recent_results} to the four most-recent of these results (WArP 2.3l, ZEPLIN-III, XENON10, XENON100), describing the detectors, science runs, and main results.  Later, in Section \ref{sec:light_WIMP} I focus on low-mass WIMPs ($\sim$10\,GeV/c$^2$) that have recently been a hot topic in the field, and describe what sensitivity noble liquids can provide to this type of dark matter candidate, in comparison with apparent detections in two non-noble-liquid experiments.

\section{Recent Dark Matter Results with Noble Liquid Detectors}
\label{sec:recent_results}

The four most-recent dark matter results from noble liquid detectors all use a detector design known as a dual-phase time projection chamber (TPC).  These detectors detect both the scintillation photons and electrons emitted from an interaction site in the LAr or LXe.  Photomultiplier tubes (PMTs), instrumented above and/or below the active liquid volume, detect a fraction of the scintillation photons.  An applied static electric field drifts electrons away from the interaction site and up to the liquid surface, where a separate electric field extracts these electrons into the gas where they collide with gas atoms and stimulate further scintillation as they travel to the anode.  This additional scintillation signal (caused by the electrons traveling through the gas) is known as \emph{proportional scintillation light}, as its intensity is proportional to the number of extracted electrons, and is also detected by the PMTs.  Therefore, the PMTs are responsible for detecting both scintillation (`S1') and charge (`S2').  The electrons drift through the liquid at a constant velocity, and therefore the $z$-position of the event is given by the delay time between S1 and S2.  Additionally, because the S2 signal is emitted always directly below the top PMTs, the pattern of detected photons in this signal can be used to determine the $x,y$-position as well.

As the ionization density along a track from a nuclear recoil is generally much higher than that from an electronic recoil, the efficiency for electrons to recombine with parent ions is much higher in nuclear recoils.  Therefore, the ratio of S2 to S1 is used as a parameter to distinguish the two types of recoils.  This parameter by itself provides electromagnetic background rejection at the level of 99.0\% to 99.9\%.  Additionally, the time structure of the scintillation emission can be characterized as a combination of a fast ($\sim$few ns) singlet de-excitation and a slow triplet de-excitation; the ratio of the intensity of scintillation from the fast and slow components can be used as an additional parameter to distinguish between electronic and nuclear recoils.  The slow component in LXe is on the order of 10s of ns, and therefore does not provide a very good discrimination parameter, particularly at the low energies of interest in such a dark matter search.  However, in LAr the slow scintillation component is roughly three orders of magnitude slower than the fast component, and allows this additional discrimination parameter to reject electronic recoils at the level of $\sim$99.99999\% efficiency.

\subsection{WArP 2.3l}
The WArP 2.3l experiment\,\cite{Benetti:2007cd} operated a 2.3\,l LAr dual-phase TPC at the underground Laboratori Nazionali del Gran Sasso (LNGS), in central Italy.  This detector served as a prototype for the larger 100\,l LAr TPC currently running in the same location.  The active LAr volume was viewed from above by 12 PMTs, which (as described above) were responsible for measuring both the S1 and S2 signals.  Nuclear recoil discrimination is performed based on two parameters.  One parameter is log$_{10}$(S2/S1), and another parameter, $F$, quantifies the pulse shape.

\begin{figure}[htp]
	\begin{center}
		\includegraphics[width=.9\textwidth]{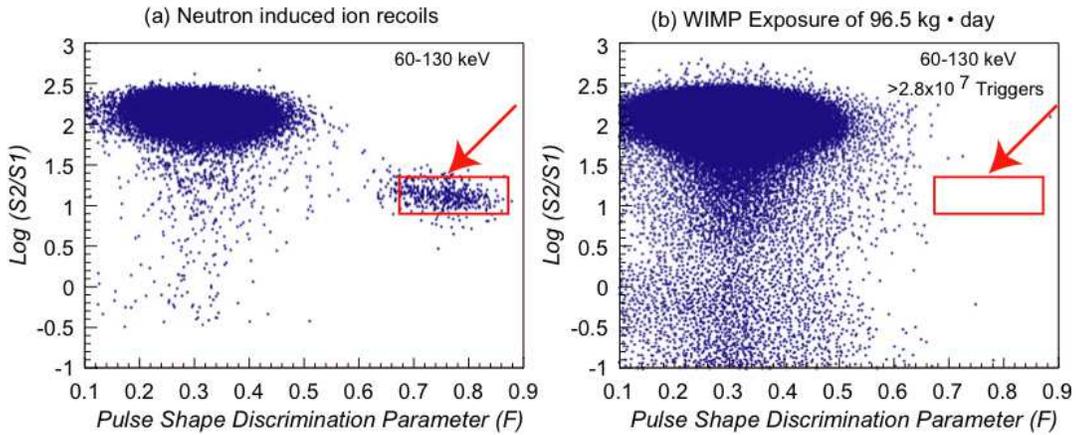}
	\end{center}
	\caption{WArP 2.3\,l events in one energy bin distributed according to the two discrimination parameters for neutron calibration data (\emph{left}) and WIMP-search data (\emph{right}).  The red box, defined based on the neutron calibration, indicates the location in this parameter space where nuclear recoils are expected, and is therefore defined to be the signal region.  Both figures taken from Ref. 9.}
	\label{fig:WARP_data}
\end{figure}

Data were collected from a central 3.2\,kg region of the detector for a total effective exposure of 100\,kg\,d.  The two discrimination parameters are calculated for each event, with the WIMP signal window defined based on calibration with a neutron source.  Figure \ref{fig:WARP_data} shows the distribution of these two discrimination parameters for one energy bin for a calibration with a neutron source (\emph{left}) and from WIMP-search data (\emph{right}).  The signal from nuclear recoils, defined on the neutron calibration data, is indicated by the red box.  In addition to this energy bin, no events were seen in the signal region for energies above 55\,keV.  This lack of events translates to an upper limit on the WIMP-nucleon scalar cross-section of $\sim$10$^{-42}$\,cm$^2$ for 100\,GeV/c$^2$ WIMPs.  The full exclusion curve is shown in Figure \ref{fig:ALL_limits}.

\subsection{ZEPLIN-III}

The ZEPLIN-III experiment\,\cite{Lebedenko:2008gb} uses a dual-phase LXe TPC, operated at the Palmer Underground Laboratory in Boulby, UK.  The detector is designed for high electric drift field (to improve nuclear recoil discrimination) and precise $x,y$-position reconstruction.  Thirty-one 2-inch PMTs view the active LXe volume from below, which is 19\,cm in diameter and 3.5\,cm in thickness.  As mentioned in Section \ref{sec:intro}, the singlet and triplet de-excitation time scales are too close together in LXe to be much use as a discrimination parameter, and therefore only the ratio of scintillation to charge can be used.

\begin{figure}[htp]
	\begin{center}
		\includegraphics[width=.5\textwidth]{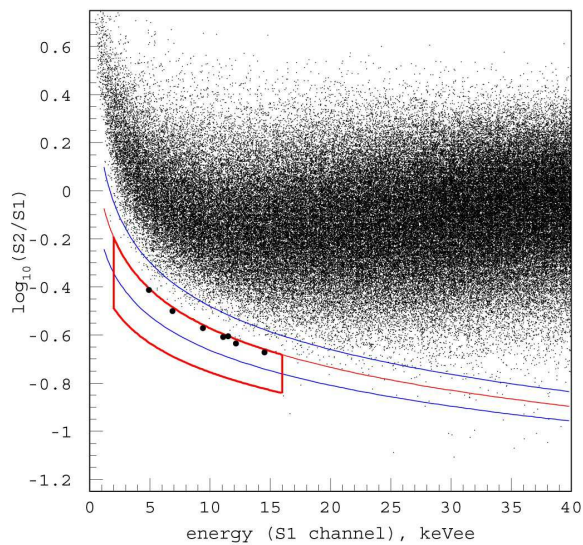}
	\end{center}
	\caption{The ZEPLIN-III nuclear recoil discrimination parameter, log$_{10}$(S2/S1), versus energy for the WIMP search data.  The thin red line indicates the centroid of the nuclear recoil distribution, its 1$\sigma$ spread shown by the thin blue lines.  The signal acceptance region, defined as the space between the centroid and the --2$\sigma$ contours, is indicated by the thick red box.}
	\label{fig:ZEPLIN_discrim}
\end{figure}

Figure \ref{fig:ZEPLIN_discrim} shows the LXe discrimination parameter, log$_{10}$(S2/S1), as a function of energy.  The units of the horizontal scale are given as ``keVee'', to indicate that these energies are reconstructed using the ``electronic-equivalent'' energy scale.  The relation between S1 and the deposited energy differs for electronic and nuclear recoils.  Seven events are seen in the signal acceptance region, following an effective exposure of 128\,kg\,d, with expected background of $11.6\pm3.0$ events from electronic recoils, and $1.2\pm0.6$ events from neutrons.  This leads to an upper limit on the WIMP-nucleon scalar interaction cross-section of $8.1\times10^{-44}$\,cm$^2$ at 60\,GeV/c$^2$.  The ZEPLIN-III detector is currently running with new PMTs and an overall electromagnetic background level $\sim$10 times lower than the data shown here.

\subsection{XENON10}

The XENON10 experiment\,\cite{Angle:2007uj} operated at the same underground facility as WArP in central Italy.  Like the ZEPLIN-III experiment, it used a dual-phase LXe TPC.  The active region was 20\,cm diameter by 15\,cm height, and viewed from above and below by 88 PMTs.  Event position reconstruction featured resolution at the level of mm for all three spatial dimensions.

\begin{figure}[htp]
	\begin{center}
		\includegraphics[width=.5\textwidth]{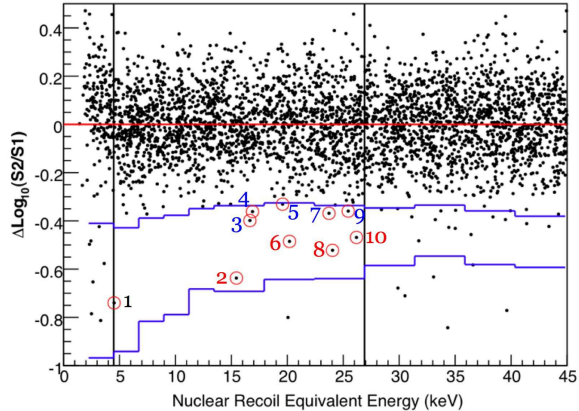}
	\end{center}
	\caption{WIMP search data from the XENON10 experiment.  Ten events are observed in the signal region, indicated by the blue lines.}
	\label{fig:XENON_discrim}
\end{figure}

Figure \ref{fig:XENON_discrim} shows the main WIMP search data of the XENON10 experiment in the same parameter spaced used in ZEPLIN-III, after an effective exposure of 136\,kg\,d.  Ten events appear in the signal acceptance region, lying between the two blue lines.  The expected background due to statistical leakage of the electronic recoils (black points) into the signal region was $7.0^{+1.4}_{-1.0}$, and is believed the be the origin of the five events labeled with blue numbers (3, 4, 5, 7, 9).  Event 1 is a noise event, and events 2, 6, 8, and 10 are consistent with a separate class of background; see Ref(\cite{Angle:2007uj}) for a description.  Nevertheless, all ten events were treated in the analysis as signal (without background subtraction), which leads to an upper limit on the scalar WIMP-nucleon cross-section of $5.5\times10^{-44}$\,cm$^2$ at 35\,GeV/c$^2$.

\subsection{XENON100}

\begin{figure}[htp]
	\begin{center}
		\raisebox{1cm}{\includegraphics[width=.5\textwidth]{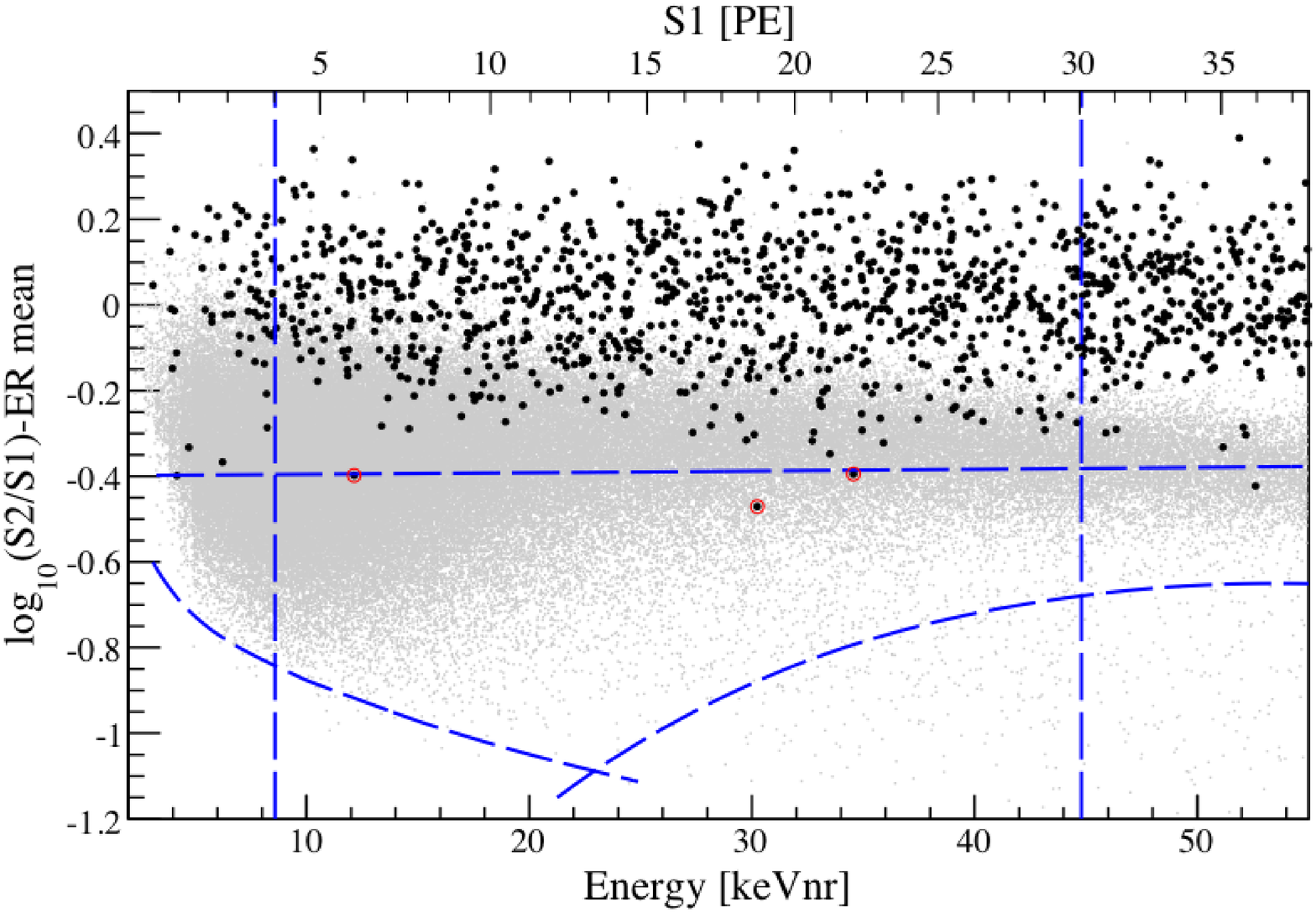}}
		\includegraphics[width=.45\textwidth]{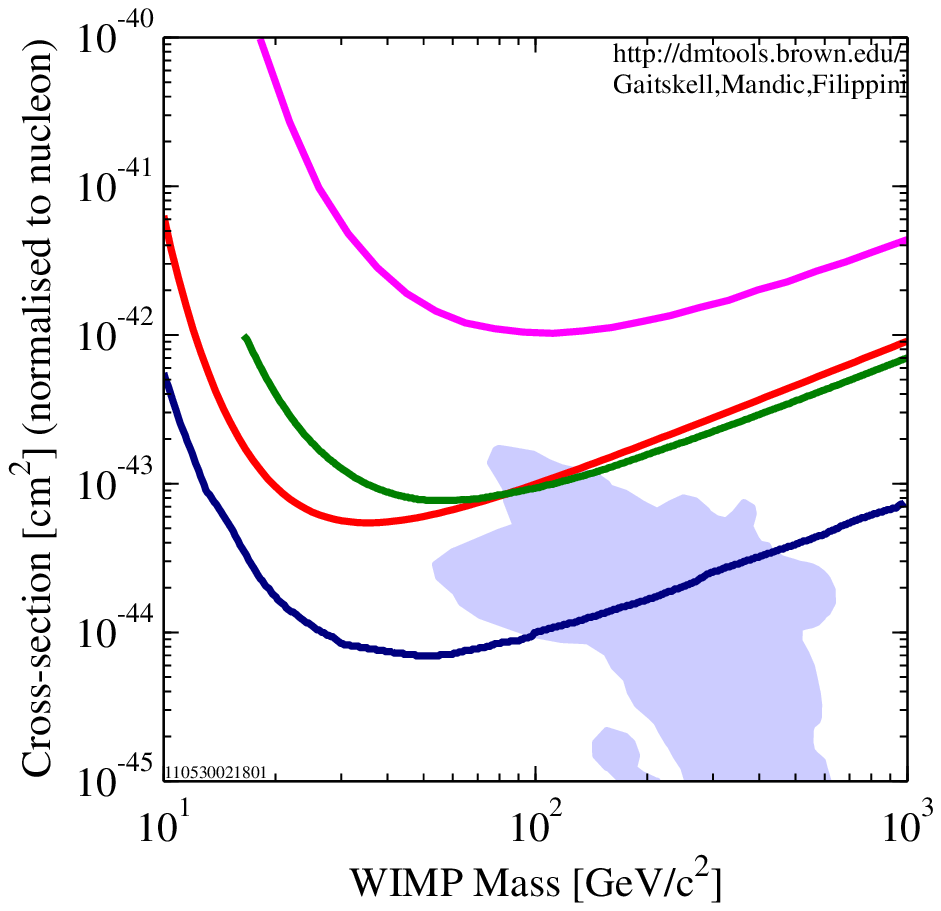}
	\end{center}
	\caption{(\emph{Left}) WIMP search data for the XENON100 experiment (black points); the nuclear recoil band, defined by a neutron calibration, is indicated by the gray points.  The signal window is the region enclosed by the blue dashed curves.  Six events existed in the signal region after unblinding, three of which were removed by an additional post-unblinding cut.  The remaining three signal events are highlighted by red circles. (\emph{Right}) Exclusion limits on the WIMP-nucleon scalar interaction cross-section as a function of WIMP mass for the four experiments discussed in the text: WArP 2.3\,l (magenta), ZEPLIN-III (dark green), XENON10 (red), XENON100 (dark blue).  Also shown is a region of parameter space favorable for the neutralino from SUSY in one calculation of CMSSM.  Plot made using DM Tools.  See text for citations.}
	\label{fig:ALL_limits}
\end{figure}

The XENON100 experiment\,\cite{Aprile:2010um} operates a very similar detector to XENON10 (and is located in the same shield where XENON10 operated), but with a detection volume larger by an order of magnitude.  The active LXe volume, 30\,cm in diameter and 30\,cm in height, is viewed from the top and bottom by 178 PMTs.  In addition to its larger size as compared with XENON10, additional steps were taken during construction to choose detector materials that are low in common radioactive contaminants (U, Th, K).  Similar to ZEPLIN-III and XENON10, the WIMP search data in XENON100 was treated by constructing a plot of the nuclear recoil discrimination parameter versus energy.  A blind analysis was performed on an effective exposure of $\sim$1.5\,tonne\,d, after which 6 events were observed in the signal acceptance region.  After unblinding, it was seen that a population of background events, arising from electronic noise, was contaminating the data.  An additional, pos-unblinding cut was constructed to target these events specifically, which resulted in 3 remaining events in the signal region, shown in Figure \ref{fig:ALL_limits} (\emph{left}).  The resulting upper limit on WIMP-nucleon scalar cross-section is $7\times10^{-45}$\,cm$^2$ at 50\,GeV/c$^2$.

The full exclusion curves (90\% C.L.) of the four experiments discussed in this section, for the WIMP-nucleon scalar cross-section, are shown in Figure \ref{fig:ALL_limits} (\emph{right}), as a function of WIMP mass.  Also shown is a recent calculation of a favored region of this parameter space by CMSSM\,\cite{Roszkowski:2007fd}.  Figure \ref{fig:ALL_limits} (\emph{right}) was made using DM Tools\,\cite{DMtools}.

\section{Light WIMPs}
\label{sec:light_WIMP}
Recently, considerable attention has been placed on the possible existence of low-mass WIMPs, of order $\sim$10\,GeV/c$^2$.  The source of this excitement comes from two experimental results that point to such a particle.  There is the long-standing observation of an annual modulation in the low-energy background rate of the DAMA/NaI and DAMA/LIBRA experiments\,\cite{Bernabei:2010mq}.  Operating NaI scintillating crystals at LNGS, DAMA/LIBRA is unable to distinguish nuclear recoils from electronic recoils, and instead uses the observed annual modulation as a dark matter identification feature.  This signal can be interpreted as being the result of WIMPs scattering off of sodium nuclei, which would imply the existence of a WIMP with mass in the range $\sim$8--12\,GeV/c$^2$, with a WIMP-nucleon scalar cross-section of $\sim$few$\times10^{-40}$\,cm$^2$.  However, it should be noted that the observed annual modulation is correlated with both the observed modulation in the cosmic muon rate\,\cite{Selvi_LVD} and the rate of ambient fast neutrons\,\cite{Ralston:2010bd} in LNGS.

The second source of excitement over low-mass WIMPs comes from the observed exponentially-falling spectrum in the background data of the CoGeNT experiment\,\cite{Aalseth:2010vx} below the normal threshold used for previous analyses.  This detector also features no discrimination between electronic and nuclear recoils, but offers the ability to reject surface events with energies above $\sim$2\,keV.  The exponential fall in the differential spectrum, extending from roughly 0.5\,keV to 1\,keV, can be fit by the expected recoil spectrum of low-mass WIMP, with mass and cross-section similar to the interpretation of the DAMA signal mentioned in the previous paragraph.  However, as many experiments should be sensitive to these regions of parameter space, the low-mass WIMP interpretation of these two signals is treated with varying degrees of skepticism and optimism within the field.

The exclusion curves shown in Figure \ref{fig:ALL_limits} (\emph{right}) are all calculated by using the scintillation signal (S1) to reconstruct the energy of each event.  The nonlinear relationship between the average scintillation signal and the energy of the recoiling nucleus is quantified by the parameter $\mathcal{L}_{\mathrm{eff}}$ (the ``effective Lindhard parameter'').  For energies of interest to the WArP result, the LAr $\mathcal{L}_{\mathrm{eff}}$ measured values show little energy dependence, and are in agreement.  However, this is very much not the case for $\mathcal{L}_{\mathrm{eff}}$ in LXe\,\cite{Manalaysay:2010mb}.  As a result, the exclusion curves for LXe contain a degree of uncertainty, and this uncertainty has lead some to question the robustness of these upper limits with respect to the possible signal detection by DAMA and CoGeNT.

\begin{figure}[htp!]
	\begin{center}
		\includegraphics[width=.6\textwidth]{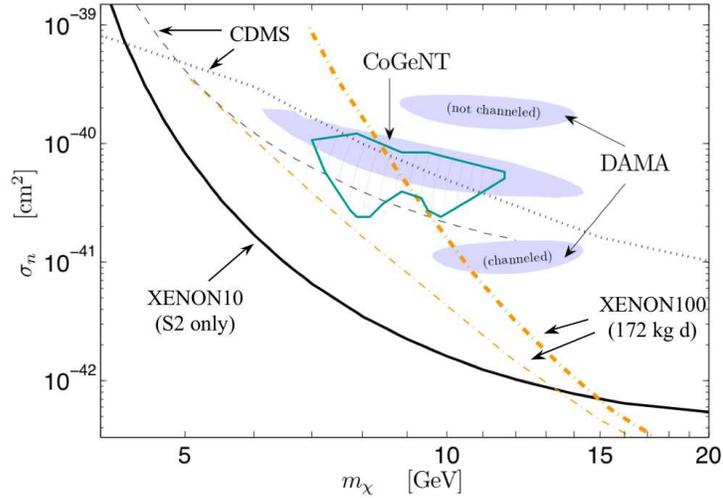}
	\end{center}
	\caption{The typical WIMP parameter space, in WIMP-nucleon scalar cross-section versus WIMP mass, focusing on the regions around 10\,GeV/c$^2$.  Regions consistent with a light WIMP interpretation of DAMA and CoGeNT are indicated by the blue regions.  Exclusion limits from several experiments are also shown (see text for explanation).}
	\label{fig:XENON10_s2_limit}
\end{figure}

Recently, however, it was pointed out that the charge signal (S2) can be used (instead of S1) to reconstruct the recoil energy of events with much greater sensitivity that what is possible with using S1 alone\,\cite{Sorensen:2010hv}.  Measurements of the charge yield, unlike $\mathcal{L}_{\mathrm{eff}}$, have shown remarkable consistency, and are additionally well matched to theoretical expectations\,\cite{Sorensen:2011bd}.  Using this insight, additional WIMP search data collected by XENON10 has been analyzed, specifically targeting low-mass WIMPs\,\cite{Angle:2011th}.  These data, not used in previous XENON10 publications, featured a reduced trigger threshold at the level of a single electron.  Using a conservative analysis threshold of 5 electrons, corresponding to 1.4\,keV, along with tight fiducial cuts, results in the exclusion curve shown in Figure \ref{fig:XENON10_s2_limit}.

The 90\% C.L.~exclusion curves in Figure \ref{fig:XENON10_s2_limit} include the low-threshold analyses of CDMS-I\,\cite{Akerib:2010pv} and CDMS-II\,\cite{Ahmed:2010wy} experiments, XENON100 exclusions (based on 172\,kg\,d)\,\cite{Aprile:2010um} using two choices of $\mathcal{L}_{\mathrm{eff}}$, and the new limit from the XENON10 S2-only analysis\,\cite{Angle:2011th}.  Regions consistent with CoGeNT (blue-green contour\,\cite{Aalseth:2010vx}) and CoGeNT and DAMA (light-blue shaded regions\,\cite{Chang:2010yk}) are also indicated.  These constraints are not weakened if one considers scattering mediated by axial vector coupling (``spin dependent'') in the case of CoGeNT.  This is because germanium and xenon both have their main spin-dependent sensitivity on couplings to neutrons, and natural Xe contains more odd isotopes than natural germanium.  Given the magnitude of this new XENON10 null result, it becomes difficult to understand the CoGeNT and DAMA signals with a light WIMP interpretation.

\end{document}